\documentclass{article}
\usepackage{amssymb}
\usepackage{amsmath}
\usepackage{amsfonts}
\usepackage{authblk}
\begin{document}
\title{Generalized action-angle coordinates in toric contact spaces}

\author{Mihai Visinescu\thanks{mvisin@theory.nipne.ro}}

\affil{Department of Theoretical Physics,

National Institute for Physics and Nuclear Engineering,

Magurele, P.O.Box M.G.-6, Romania}

\date{}
\maketitle

\begin{abstract}

In this paper we are concerned with completely integrable Hamiltonian systems in 
the setting of contact geometry. Unlike the symplectic case, contact structures are
automatically Hamiltonian. Using the Jacobi brackets defined on contact manifolds,
we discuss the commutativity of the first integrals for contact Hamiltonian systems
and introduce the generalized contact action-angle variables. We exemplify the 
general scheme in the case of the five-dimensional toric Sasaki-Einstein spaces
$T^{1,1}$ and $Y^{p,q}$.

\end{abstract}

~






\section{Introduction}

There has been considerable interest recently in contact geometry in connection with some
modern developments in mathematics and theoretical physics \cite{BG,JS}. The theory of 
contact structures is linked to many geometric backgrounds as symplectic geometry,
Riemannian and complex geometry, analysis and dynamics \cite{IVV,VV}. Contact spaces have 
shown their usefulness in gauge theories of gravity, black holes in higher dimensions, branes. 
Sasaki-Einstein manifolds whose metric cones are Calabi-Yau manifolds find applications 
in string theory in connection with AdS/CFT correspondence which relates quantum gravity
in a certain background to ordinary quantum field theory without gravity.

The isolated systems are conservative and their standard description is given in terms 
of Hamilton's equations of motion in the phase space which has a natural symplectic 
structure. Over the last decades, there have been many attempts to extend the symplectic 
Hamiltonian mechanics. For example, contact Hamiltonian mechanics is a natural candidate
for a geometric description of non-dissipative and dissipative systems \cite{BCT}. 
Contact geometries are adequate in describing mechanical systems where the Hamilton 
function explicitly depends on time. Contact Hamiltonian dynamics has been used in 
thermodynamics \cite{BL-MN} and in description of dissipative systems at the 
mesoscopic level \cite{MG}.

The aim of this paper is to investigate contact Hamiltonian dynamics for a class
of toric contact structures. In the case of toric contact spaces, the system is
completely integrable if the toric action is effective and preserve the contact 
structures. By analogy to standard symplectic dynamics, we introduce the 
action-angle variables and indicate the possibility to evaluate the frequencies
of the flow of toric action.

The paper is organized as follows: In Section 2 we review fundamentals on contact 
geometry and contact Hamiltonian dynamics. The contact dynamics is presented first 
in a coordinate-free manner and then in special local coordinates. In Section 3
we illustrate  the contact Hamiltonian dynamics through two examples related to 
toric Sasaki-Einstein spaces in five-dimensions. In Section 4 we provide some 
closing remarks.

\section{Preliminaries}

\subsection{Contact geometry}

A contact manifold $(M,\eta)$ is a $(2n+1)$-dimensional manifold $M$ 
endowed with a contact $1$-form $\eta$ such that \cite{BG}
\begin{equation}
\eta \wedge (d\eta)^n \neq 0\,.
\end{equation}

Associated with a contact form $\eta$ there exists a unique vector field
$R_\eta$ called the {\it Reeb vector field} defined by the contractions 
(interior products):
\begin{subequations}\label{Reeb}
\begin{equation}
i(R_\eta) \eta = 1\,,
\end{equation}
\begin{equation}
i(R_\eta) d\eta = 0 \,.
\end{equation}
\end{subequations}

The tangent bundle $TM$ may be decomposed into \cite{PL}
\begin{equation}\label{TM}
TM = \mathbb{R} R_\eta \oplus \mathcal{H} \,,
\end{equation}
and by duality we have for the cotangent bundle $T^* M$
\begin{equation}
T^*M = \mathbb{R} \eta \oplus \mathcal{K} \,.
\end{equation}

The subbundle $\mathcal{H}$ (or the {\it horizontal distribution}), of rank $2n$, is
the kernel of $\eta$. $\mathcal{K}$ is the annihilator of $\text{ker}\,d \eta =
\mathbb{R} R_\eta$ and its sections are called {\it semi-basic forms} satisfying
the relation
\begin{equation}
i(R_\eta) \varphi = 0\,.
\end{equation}

According to \eqref{TM} every vector field $X$ on $M$ may be decomposed as
\begin{equation}\label{dec}
X = (i(X) \eta) R_\eta + \hat{X}
\end{equation}
where $\hat{X}$ is the horizontal part of $X$. Every $1$-form $\psi$ may be in turn
decomposed as
\begin{equation}
\psi = (i(R_\eta)\psi)\eta + \hat{\psi}\,,
\end{equation}
where $\hat{\psi}$ is the semi-basic component of $\psi$.

The mapping
\begin{equation}
\eta^\flat : X \mapsto -i(X) d\eta\,,
\end{equation}
carries any vector field $X$ on $M$ into a semi-basic form. We will denote the 
inverse isomorphism of $\eta^\flat$ by $\eta^\sharp$.

A vector field $X$ on $(M,\eta)$ is an infinitesimal contact automorphism
if and only if there exists a differentiable function $\rho$ such that
\begin{equation}\label{ia}
\mathcal{L}(X) \eta = \rho \eta\,.
\end{equation}

In what follows we shall write eq. \eqref{dec} in the form
\begin{equation}\label{Xf}
X_f = f R_\eta + \hat {X}_f \,,
\end{equation}
where $fR_\eta$ and $\hat {X}_f$ are, respectively, the vertical and horizontal
components with
\begin{equation}\label{f}
f = i(X_f) \eta \,.
\end{equation}

With the help of Cartan's formula connecting the Lie derivative with the interior 
product, $\mathcal{L}(X) = d \circ i(X) + i(X) \circ d $, eq. \eqref{ia} may be written
\begin{equation}\label{frho}
d f + i(X_f) d \eta = \rho \eta \,.
\end{equation}
Using the properties \eqref{Reeb} of the contact form $\eta$ we have
\begin{equation}\label{rho}
\rho = i(R_\eta) d f\,.
\end{equation}
The condition $\rho = 0$ expresses the fact that $f$ is a {\it first integral} 
of the vector field $R_\eta$ being a constant along the flow of the vector 
field $R_\eta$.

A chosen contact form $\eta$ on $M$ defines an isomorphism $\Phi$ from the vector space
of infinitesimal contact automorphisms onto the set $C^\infty (M)$ of smooth
functions on $M$:
\begin{equation}\label{Phi}
\Phi(X_f) = f = i(X_f) \eta \,,
\end{equation}
with the inverse
\begin{equation}
\Phi^{-1}(f) = f R_\eta + \eta^\sharp (d f - (i(R_\eta) d f)\eta)\,.
\end{equation}
Let us remark that the Reeb vector field $R_\eta = \Phi^{-1}(1)$ is an infinitesimal 
automorphism of the contact form $\eta$ with $\rho = 0$.

\subsection{Contact Hamiltonian systems}
On a symplectic manifold $(M,\Omega)$, $\Omega$ being the symplectic form, the system 
of the 1-st order differential equations
\begin{equation}\label{HE}
\dot{x} = X_H \,,
\end{equation}
for some smooth function $H$ on $M$ is called a Hamiltonian system.

The development of the theory of completely integrable systems in contact geometry has 
been more recent starting with the influential work of Banyaga and Molino \cite{BM}.

In the frame of contact geometry, the vector field $X_f = \Phi^{-1} (f)$ \eqref{Phi} is 
called the contact Hamiltonian vector field and the analog of \eqref{HE} 
\begin{equation}\label{HXf}
\dot{x} = X_f \,,
\end{equation}
represents the {\it contact Hamiltonian equation} corresponding to $f$.
Taking into account \eqref{rho}
we get that $X_f$ is an infinitesimal automorphism of $\eta$ if and only if $df$ is 
semi-basic \cite{BJ}. 

It is often convenient to consider the Reeb vector field $R_\eta$ 
as the Hamiltonian vector field with $1=\eta (R_\eta)$ as the Hamiltonian. In this case 
the Hamiltonian contact structure is said to be of {\it Reeb type} and the 
Hamiltonian is understood to be the constant function $1$.

In connection with the isomorphism $\Phi$ \eqref{Phi}, the Lie algebra structure of 
$C^\infty (M)$ is given by the Jacobi bracket \cite{LM,CB}
\begin{equation}
\begin{split}
[f,g]_\eta &= \Phi[X_f,X_g]\\
&= - i(X_g) df + f i(R_\eta) dg\\
&= -i(X_f) i(X_g) d\eta + f i(R_\eta) dg - g i(R_\eta) df \,. 
\end{split} 
\end{equation}

Assuming that $f$ and $g$ are first integrals of the vector field $R_\eta$ we have
\begin{equation}
[f,g]_\eta = d\eta (X_f, X_g) \,.
\end{equation}

A Hamiltonian contact structure of Reeb type is said to be {\it completely integrable}
if there exists $(n+1)$ first integrals $f_0=1, f_1,\dots , f_n$ that are independent 
and in involution. In addition a completely integrable contact Hamiltonian system is 
said to be of toric type if the corresponding vector fields $X_{f_0}= R_\eta, X_{f_1},
\dots, X_{f_n}$ form the Lie algebra of a torus $T^{n+1}$. The action of a torus $T^{n+1}$
on a contact $(2n+1)$-dimensional manifold $(M,\eta)$ is completely integrable if it is 
effective and preserve the contact structure $\eta$ \cite{EL}.

\subsection{Formulae in local coordinates}

In view of the concrete examples which will be studied in the next Sections, it is convenient
to write some of the above formulae in local coordinates.

Let us consider in a neighborhood $U$ of a point $x$ of $M$ an adapted system of local 
coordinates $(x^0,x^1,\dots,x^n,y^1,\dots,y^n)$.  According to Darboux's theorem, in the case 
of contact geometry, the contact form can be written as
\begin{equation}\label{etac}
\eta = d x^0 - \sum_{k=1}^n y^k dx^k\,,
\end{equation}
and the Reeb vector field defined by $\eta$ is
\begin{equation}\label{Reebc}
R_\eta = \frac{\partial}{\partial x^0} \,.
\end{equation}

In the above adapted system of local coordinates, a vector field can be written as
\begin{equation}
X = a_0 \frac{\partial}{\partial x^0} + \sum_{k=1}^n a_k\frac{\partial}{\partial x^k} 
+\sum_{k=1}^n b_k\frac{\partial}{\partial y^k} \,.
\end{equation}

According to eq. \eqref{f}, a vector field $X_f$ is associated with a function
\begin{equation}
f = a_0 - \sum_{k=1}^n a_k y^k \,,
\end{equation}
and from eq. \eqref{rho} we get that
\begin{equation}
\rho = \frac{\partial f}{\partial x^0} \,.
\end{equation}

On the other hand, from eq. \eqref{frho} we get
\begin{equation}
\begin{split}
a_k &= - \frac{\partial f}{\partial y^k} \,, \\
b_k &=  \frac{\partial f}{\partial x^k} + \frac{\partial f}{\partial x^0} y^k\,.
\end{split} 
\end{equation}
Therefore a vector field $X_f= \Phi^{-1} (f)$ has in an local system of coordinates
the form
\begin{equation}
\begin{split}
X_f &= \Bigl( f - y^k \frac{\partial f}{\partial y^k} \Bigr) 
\frac{\partial }{\partial x^0}
- \frac{\partial f}{\partial y^k} \frac{\partial}{\partial x^k} +
\Bigl(\frac{\partial f}{\partial x^k} + y^k \frac{\partial f}{\partial x^0} \Bigr) 
\frac{\partial }{\partial y^k}\\
&= f R_\eta + \eta^\sharp \biggl[ \Bigl(\frac{\partial f}{\partial x^k} + 
\frac{\partial f}{\partial x^0} y^k\Bigr) d x^k + 
\frac{\partial f}{\partial y^k} dy^k\biggr]\,.
\end{split} 
\end{equation}
Here and in the sequel, we use the convention that repeated indices are summed over.

Finally, the Jacobi bracket of two functions $f$ and $g$ may be expressed as
\begin{equation}\label{Jb}
\begin{split}
[f,g]_\eta = & \Bigl( f- y^k \frac{\partial f}{\partial y^k} \Bigr)
\frac{\partial g}{\partial x^0} - 
\Bigl( g- y^k \frac{\partial g}{\partial y^k} \Bigr)
\frac{\partial f}{\partial x^0}\\
& + \Bigl( \frac{\partial f}{\partial x^k} \frac{\partial g}{\partial y^k}-
 \frac{\partial g}{\partial x^k} \frac{\partial f}{\partial y^k} \Bigr).
\end{split} 
\end{equation}

\section{5-dimensional Sasaki-Einstein spaces}

A contact Riemannian manifold $M$ equipped with a metric $g$ is Sasakian if its 
metric cone
\begin{equation*}
(C(M), \bar{g}) = (\mathbb{R}_+ \times M, dr^2+r^2g)\,,
\end{equation*}
is K\"{a}hler. Here $r \in (0,\infty)$ may be considered as a coordinate on the
positive real line $\mathbb{R}_+$. Moreover if the Sasaki manifold is Einstein
\begin{equation*}
\text{Ric}_g = 2n g \,,
\end{equation*}
then the K\"{a}hler metric cone is Ricci flat ($\text{Ric}_{\bar{g}} =0)$, i.e. 
a Calabi-Yau manifold.

\subsection{Sasaki-Einstein space $T^{1,1}$}

One of the most familiar examples of homogeneous toric Sasaki-Einstein 
$5$-dimensional manifold is the space $T^{1,1}$. The metric on $T^{1,1}$
may be written down explicitly by utilizing the fact that it is a $U(1)$ bundle
over $S^2 \times S^2$. We choose the coordinates  
$(\theta_i, \phi_i)\,,\, i=1,2$ to parametrize the two spheres $S^2$ in the 
standard way, while the angle  $\psi \in [0, 4 \pi)$ parametrizes the $U(1)$ 
fiber. Using these parametrizations the metric on $T^{1,1}$ may be written as 
\cite{CO,MS}
\begin{equation}\label{metricT}
\begin{split}
ds^2(T^{1,1}) = & \frac16 (d \theta^2_1 + \sin^2 \theta_1 d \phi^2_1 +
d \theta^2_2 + \sin^2 \theta_2 d \phi^2_2) \\
& +\frac19 (d \psi + \cos \theta_1 d \phi_1 + \cos \theta_2 d \phi_2)^2 \,.
\end{split}
\end{equation}
In what follows we introduce $\nu = \frac12 \psi$ so that $\nu$ has canonical 
period $2\pi$.

The globally defined contact $1$-form $\eta$ is:
\begin{equation}\label{etaT}
\eta =\frac13 (2d \nu +\cos \theta _1 d \phi _1+\cos \theta _2 d \phi _2) \,,
\end{equation}
and the Reeb vector field $R_\eta$ has the form
\begin{equation}
R_\eta = \frac32 \frac{\partial}{\partial \nu} \,.
\end{equation}

We employ the basis \cite{MS} for an effectively acting $\mathbb{T}^3$ action
\begin{equation}\label{efa}
\begin{split}
\mathbf{e}_1& =\frac \partial {\partial \phi _1}+\frac 12\frac \partial {\partial \nu}
\,, \\
\mathbf{e}_2& =\frac \partial {\partial \phi _2}+\frac 12\frac \partial {\partial \nu}
\,, \\
\mathbf{e}_3& =\frac \partial {\partial \nu }\,,
\end{split}
\end{equation}
which preserves the contact structure $\eta$.

As it was explained in Section 2.2 the effective action of the torus $T^3$ 
\eqref{efa} on the Sasaki-Einstein space $T^{1,1}$ is completely integrable. 
Let $\mathcal{F} = (f_0,f_1,f_2)$ be the set of independent first integrals in involution 
and $\mathcal{X} = (R_\eta, X_{f_1}, X_{f_2})$ the corresponding set of infinitesimal
automorphisms of $\eta$. Let $T$ be a compact connected component of the level set
$\{f_1=c_1,f_2=c_2\}$ and $df_1 \wedge df_2 \neq 0$ on $T$. Then $T$ is diffeomorphic
to a $T^3$ torus. There exist a neighborhood $U$ of $T$ and a diffeomorphism 
$\phi\,:\, U \rightarrow T^3 \times D$
\begin{equation}
\phi (x) = (\vartheta_0,\vartheta_1,\vartheta_2,y_1,y_2)\,,  
\end{equation}
where $D\in \mathbb{R}^2$, such that the contact form has the following 
canonical expression \cite{BM,LB}:
\begin{equation}
\eta_0 = (\phi ^{-1})^* \eta = y_0 d\vartheta_0 +  y_1 d\vartheta_1 +  
y_2 d\vartheta_2 \,.
\end{equation}
We refer to the local coordinates $(y_i, \vartheta_i)$ as {\it generalized
contact action-angle coordinates} \cite{BJ,JJ}. Note that
$\eta_0(\frac{\partial}{\partial \vartheta_i}) = y_i$
are the contact Hamiltonians of the independent set of vector fields $\mathcal{X}$.
Let us remark that the action of the torus $T^3$ is given by translations of 
the angles $\vartheta_i$.

Taking into account the $1$-form $\eta$ \eqref{etaT} it is convenient to choose
\begin{equation}
\vartheta_0 = \frac23 \nu\,,\, \vartheta_1 = \phi_1\,,\, \vartheta_2 = \phi_2\,,
\end{equation}
and accordingly we have
\begin{equation}
y_0 = 1 \,,\, y_1= \frac 13 \cos\theta_1\,,\, y_2= \frac 13 \cos\theta_2\,.
\end{equation}

These functions are first integrals of the Hamiltonian contact structure 
\begin{equation}\label{firstint}
f_0 = y_0 \equiv 1 \, , \, f_i= y_i  = \frac 13 \cos\theta_i \quad,\quad i=1,2\,, 
\end{equation}
which are independent and in involution
\begin{equation}
[1,f_i]_\eta = [f_i,f_j]_\eta =0 \quad,\quad i,j=1,2\,, 
\end{equation}
as can be seen through a direct evaluation of the respective Jacobi brackets \eqref{Jb}.

The flows of the set $\mathcal{X}$  on invariant tori is quasi-periodic
\begin{equation}\label{flow}
(\vartheta_0,\vartheta_1,\vartheta_2)\rightarrow 
(\vartheta_0 +t\omega_0,\vartheta_1+t\omega_1,\vartheta_2+t\omega_2)\,,
\end{equation}
where the {\it frequencies} $\omega_i$ depend only on $y_i$.

In order to construct effectively the flow of $X_f$ and find the frequencies $\omega_i$
we define the family of $1$-forms
\begin{equation}\label{etat}
\eta_t = \eta_0 + t df\,,
\end{equation}
where $f$ is one of the first integrals of the Reeb vector field $R_\eta$. 
We observe that $\eta_t$ is a contact form also having the Reeb vector field $R_\eta$. 
Following \cite{BM} we consider the vector field $X= -f R_\eta$ and let $\phi_t$ 
the flow of this vector field. Because $f$ is a first integral of the $T^3$ action,
$\phi_t$ commutes with this action.
Using the Moser's deformation \cite{BM,HG} we have 
\begin{equation}
\mathcal{L}(X) \eta_t = - df = - \frac{\partial \eta_t}{\partial t} \,,
\end{equation}
which imply 
\begin{equation}
\frac{d}{d t} (\phi^*_t \eta_t) = \phi^* \Bigl( \mathcal{L}(X) \eta_t +
\frac{\partial \eta_t}{\partial t}\Bigr) =0\,.
\end{equation}

Therefore $\phi^*_1 \eta_1 = \eta_0$ and we can obtain the coordinates in which the 
$1$-form \eqref{etat} has the canonical expression. In our case choosing the first
integrals $f_i=y_i$ as in ec.\eqref{firstint}, a simple calculation permits us to 
extract the frequencies:
\begin{equation}
\omega_i  = \ln \cos\theta_i \quad,\quad i=1,2\,, 
\end{equation}

\subsection{Sasaki-Einstein space $Y^{p,q}$}

The metric of the Sasaki-Einstein space $Y^{p,q}$ is given by the line element \cite{MS}
\begin{equation}
\begin{split}\label{Ypq}
ds^2 & = \frac{1- y}{6}( d \theta^2 + \sin^2 \theta\, d \phi^2)
+  \frac{1}{w(y)q(y)} dy^2
+ \frac{q(y)}{9} ( d\psi - \cos \theta \, d \phi)^2\\
& \quad  +
w(y)\left[ d\alpha + \frac{a -2y+  y^2}{6(a-y^2)}
[d\psi - \cos\theta \, d\phi]\right]^2\,,
\end{split}
\end{equation}
where
\begin{equation}
\begin{split}
w(y) & = \frac{2(a-y^2)}{1-y}\,,\\
q(y) & = \frac{a-3y^2 + 2 y^3}{a-y^2}\,.
\end{split}
\end{equation}

A detailed analysis  of the metric $Y^{p,q}$ \cite{GMSW} showed that 
it is globally well-defined and there are a countable infinite number of 
Sasaki-Einstein manifolds characterized by two relatively prime positive integers
$p, q$ with $p<q$. For $0< a<1$ one can take the range of the angular coordinates 
$(\theta, \phi, \psi)$ to be $ 0\leq \theta \leq \pi$, $ 0\leq \phi  \leq 2\pi$, 
while $y$ lies between the negative and the smallest positive zeros of $q(y)$. 
Finally, the period of $\alpha$ is chosen so as to describe a principal $S^1$ bundle
over $B_4 = S^2 \times S^2$. For any $p$ and $q$ coprime, the space $Y^{p,q}$ is
topologically $S^2 \times S^3$ and one may take \cite{MS,GMSW}
\begin{equation}
0 \leq \alpha \leq 2 \pi \ell\,,
\end{equation}
where
\begin{equation}
\ell = \frac{q}{ 3q^2 - 2 p^2 + p(4 p^2 - 3 q^2 )^{1/2}}\,.
\end{equation}

The contact $1$-form $\eta$ is \cite{MS,MVMPLA}
\begin{equation}\label{etaY}
\eta = -2y d\alpha + \frac{1-y}3 (d\psi - \cos\theta d\phi)\,,
\end{equation}
and the Reeb vector field is 
\begin{equation}\label{ReebY}
R_{\eta} = 3 \frac {\partial}{\partial \psi}
- \frac{1}{2}\frac {\partial}{\partial \alpha}\,.
\end{equation}

The orbits of the Reeb vector field \eqref{ReebY} may or may not close. The geometries 
$Y^{p,q}$ with $4p^2 - 3 q^2$ a square are examples of quasi-regular manifolds for which
the orbits of the Reeb vector field close corresponding to a locally free $U(1)$ action on
$Y^{p,q}$. On the other hand, for $4p^2 - 3 q^2$  not a square the orbits of the Reeb 
vector field do not close generating an action $\mathbb{R}$ on $Y^{p,q}$, with the orbits 
densely filling the orbits of a torus and in this case the Sasaki-Einstein manifold is said 
to be irregular.

In relation to the angular variables $\phi,\psi,\alpha$, the basis for an
effectively acting
$\mathbb{T}^3$ action is \cite{MS,SVVAP}
\begin{equation}\label{be}
\begin{split}
\mathbf{e}_1& =\frac{\partial}{\partial \phi}+\frac{\partial}{\partial \psi}\,, \\
\mathbf{e}_2& =\frac{\partial}{\partial \phi} -\frac{(p-q)\ell}{2}\frac{\partial} {\partial
\alpha}\,, \\
\mathbf{e}_3& =\ell\frac{\partial}{\partial \alpha}\,.
\end{split}
\end{equation}

In order to put the contact form \eqref{etaY} and the Reeb vector field \eqref{ReebY} in the
canonical forms \eqref{etac} and \eqref{Reebc} we introduce the angle variables
\begin{equation}
\vartheta_0 = \frac{\psi}{3}\,,\, \vartheta_1 = -6\alpha - \psi\,,\, \vartheta_2 = \phi\,,
\end{equation}
and the generalized action variables
\begin{equation}
y_0 \equiv 1 \,,\, y_1= \frac{y}{3}\,,\, y_2= \frac{y-1}{3}\cos\theta\,.
\end{equation}
These functions are first integrals of the Hamiltonian contact structure, independent and 
in involution. 

The corresponding set of infinitesimal automorphisms is $\mathcal{X} = 
(R_\eta, X_{y_1}, X_{y_2})$. The flows of the set $\mathcal{X}$ on invariant tori is 
quasi-periodic and the evaluation of the frequencies proceeds as in the case of the
space $T^{1,1}$.

\section{Concluding remarks}

An important point of interest in physics is to find the conserved quantities and
investigate the integrability of the systems. Having in mind that Sasaki-Einstein 
manifolds have become of significant interest in many areas of physics, we investigated 
the integrability in the frame of contact geometry. For example, in string theory, 
in connection with AdS/CFT correspondence, a larger class models consists of type 
$IIB$ string theory on the background $AdS_5 \times M_5$ with $M_5$ a five-dimensional 
Sasaki-Einstein space.

The analyses of integrability of geodesics of the five-dimensional Sasaki-Einstein spaces 
$T^{1,1}$ \cite{SVVEPL,BV} and $Y^{p,q}$ \cite{SVVAP,BV} show that the geodesic 
motions in these spaces are completely integrable. The description of the integrability 
of geodesic in terms of action-angle variables gives a comprehensive understanding of
dynamics \cite{MV,MVPTEP}. The presence of resonant frequencies gives way to chaotic 
behavior when the integrable Hamiltonian is perturbed by a small non-integrable piece. 
The action-angle approach offers strong support for the observation that certain 
classical string configurations in $AdS_5 \times M_5$ with $M_5$ in a large class of 
Einstein spaces is non-integrable \cite{BZ1}.

In the present paper we move the analysis of integrability from the ten-dimensional 
phase space for the geodesic motions in Sasaki-Einstein spaces to the integrability in
five-dimensional contact geometry. Unlike the symplectic case, the contact structures 
are automatically Hamiltonian. Moreover, for the $Y^{p,q}$ and $T^{1,1}$ toric 
manifolds, the torus action $T^3$ is effective and preserve the contact structure 
implying the complete integrability.

It is possible to introduce generalized action-angle variables which are similar to 
the ones in Hamiltonian dynamics. However the explicit construction of action-angle 
variables, as in the case of symplectic geometry, is more less equivalent
to solving the equations. This is a quite difficult task and a general method does not 
exist, even the equations are integrable. In particular we used the Moser's deformation 
of the contact forms and analyzed the flow of the Hamiltonian contact vector fields.

Motivated by the recent applications of contact geometry in some physical problems, 
the contact integrability deserves further studies. It would be interesting to extend 
the action-angle formulation for a better understanding of time-dependent and dissipative 
Hamiltonian systems.

\section*{Acknowledgments}

The author is indebted to B. Jovanovi\'{c} for correspondence.
This work has been supported by the project {\it NUCLEU 18 09 01 01/2018}.

\end{document}